\documentclass[preprint,showpacs,showkeys]{revtex4}

\usepackage{graphicx}
\usepackage{dcolumn}
\usepackage{bm}
\usepackage[colorlinks]{hyperref}

\begin{document}


\title{Emergence of an Onion-like Network in 
Surface Growth \\
and Its Strong Robustness
}

\author{Yukio Hayashi}
\email{yhayashi@jaist.ac.jp}
\author{Yuki Tanaka}
\affiliation{
Japan Advanced Institute of Science and Technology,\\
Ishikawa, 923-1292, Japan}

\date{\today}

\begin{abstract}
We numerically investigate that optimal robust onion-like networks 
can emerge even with the constraint of surface growth in supposing 
a spatially embedded transportation or communication system. 
To be onion-like, moderately long links are necessary in the 
attachment through intermediations 
inspired from a social organization theory.
\end{abstract}



\maketitle


\section{Introduction}
It was found that 
onion-like networks with positive degree-degree correlations 
have the optimal tolerance of connectivity against 
attacks \cite{Tanizawa12,Schneider11} 
in the state-of-the-art Network Science (at a boundary area of physics 
and computer science). 
In the network, onion-like structure is visualized 
by links between similar degree nodes because of the positive correlations, 
when nodes are set on concentric circles arranged 
in decreasing order of degrees from core to peripheral. 
Unfortunately, 
real technological networks (Internet or World-Wide-Web) 
are vulnerable with negative correlations 
\cite{Newman03}.
While there exist rewiring methods \cite{Schneider11,Wu11} 
for a network in order to enhance the correlations positively 
and to be onion-like, they are not realistic.
Because many already connected relations are discarded. 

Recently, incrementally growing methods have been proposed 
for constructing an onion-like network \cite{Hayashi14,Hayashi18a}. 
One is based on copying and adding shortcut links \cite{Hayashi14}. 
Another is based on intermediations \cite{Hayashi18a} inspired from 
some case studies in an organization theory \cite{Nishiguchi07}, 
in which the importance of long-distance relations has been suggested 
for both robustness of connectivity and efficiency of path 
in the networks such as supply chain and business collaboration. 
The established connections via intermediations probably work well 
for managing cross-border operations. 
The copying \cite{Hayashi14} is a modification of the duplication-divergence 
model \cite{Sole02} inspired from a biologically growing mechanism. 
Moreover, it has been shown that 
onion-like networks are obtained 
even with a constraint of surface growth in the method based on 
copying and adding shortcut links \cite{Hayashi16}. 
By the constraint, 
the position of new node is limited on surface, however such a 
spatial embedding is rather natural 
such as in transportation or communication systems. 
In this paper, we investigate whether or not an onion-like network 
emerges in the growing method based on intermediations, when it is 
spatially embedded with a constraint of surface growth.

\section{How to construct a network}
\subsection{Incrementally growing method}
We briefly survey the incrementally growing methods 
based on intermediations (MED) \cite{Hayashi18a,Hayashi18b}
for constructing onion-like networks.
From an initial configuration, 
e.g. a complete graph $K_{m}$ of $m$ nodes, 
a new node is added and connect to existing nodes with even number 
$m$ links at each time step. 
In the methods, 
a pair of attachments is repeatedly performed for the number of links.
In other words, 
the destination nodes of $m/2$ links are decided by one rule: 
random attachment \cite{Barabasi99}, 
and the remaining them of $m/2$ links 
are decided by another rule: attachment by MED. 
Either of the following MED-kmin and MED-rand is applied, 
the number of links emanated from a new node 
is $m/2 + m/2 = m$ at each time step.

\begin{description}
  \item[MED-kmin] One of the pair is chosen uniformly at random (u.a.r) 
    from the existing nodes by random attachment. 
    As another of it, a node with the minimum degree is chosen by 
    intermediations in $\mu$ hops form its randomly chosen pair node. 
    Intermediations in $\mu$ hops 
    mean attachments to the $(\mu + 1)$-th neighbors. 
    $\mu \geq 0$ is a small integer.
  \item[MED-rand] One of the pair is the same as MED-kmin.
    For another of it, 
    instead of selecting the node with the minimum degree in MED-kmin, 
    a node is chosen u.a.r in the $(\mu + 1)$-th neighbors 
    from its randomly chosen pair node. 
\end{description}
For prohibited multi-links, choosing other node is tried. 
Intuitively, 
random attachment contributes to enhance the correlations between 
large degree nodes, since older nodes tend to get more links by 
random attachment \cite{Barabasi99} and to connect each other. 
While another attachment by MED
contributes to enhance the correlations between small degree nodes, 
since it often happens that 
the degree of the $(\mu + 1)$-th neighbor is small and 
the degree $m$ of new node is the minimum in the network. 
However, these roles are exchanged 
in the case with a constraint of surface growth as mentioned later. 

\subsection{Constraint of surface growth}
Some most important classes of surface growth include 
{\it Diffusion-Limited Aggregation} (DLA) \cite{Witten83}, 
{\it Invasion Percolation} (IP) \cite{Wilkinson83}, and 
{\it Eden growth} \cite{Eden61}, 
which can be used as a basis for understanding a wide range of 
pattern-formation phenomena with (e.g. fractal) disorderly growth.
Thus, 
as similar to the copying network \cite{Hayashi16}, 
we consider the typical diffusive growth on surface, 
whose position of new node is determined by DLA, IP, 
or Eden model on a square lattice. 

\begin{description}
  \item[\underline{DLA model}] It means 
    a dendric extension in technological or social network system. 
  \item[\underline{IP model}] It means an extension with some holes 
    in avoiding geographical obstacles (e.g. mountain or rivers)
    for the network construction.
 \item[\underline{Eden model}] 
    It imitates a cell division process, 
    which means the most local extension to neighbor space
    in a simple way.
\end{description}

\begin{figure}[htb]
  \centering
  \includegraphics[width=0.73\linewidth]{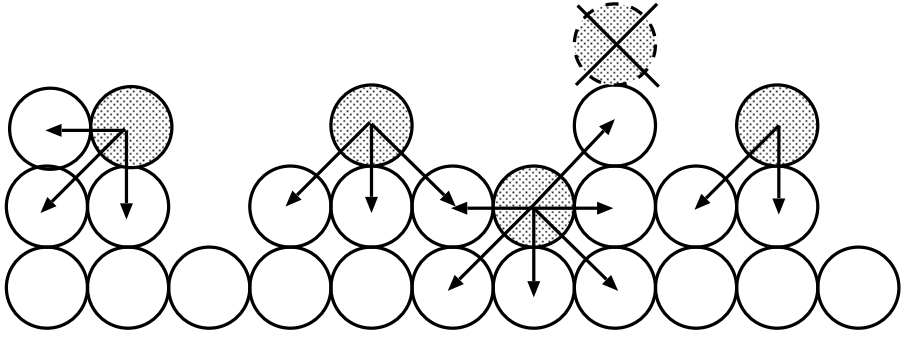}
\caption{Candidate positions of new node on surface.}
\label{fig_attach_surface}
\end{figure}

After determining the position of new node by DLA, IP, or Eden model, 
it connect to the existing nodes 
as shown in Fig. \ref{fig_attach_surface}. 
Showing the rotated cases by $90^{\circ}$, $180^{\circ}$, and $270^{\circ}$ 
are omitted because of the same. 
The shaded circles and arrows denote candidates of the position of 
new node and its links, while the dashed circle denotes a prohibited case 
because of lack of contact nodes less than $m/2$. 
The open circles denote already existing nodes in the network. 
The contact of new node to surface corresponds to a random attachment, 
since the position of new node on surface 
is randomly chosen according to DLA, IP, or Eden model. 
As random attachment on surface, 
the destinations of 
$m/2$ links are chosen from the neighbors (at least 2 and at most 7) 
indicated by arrowheads. 
We set $m=4$, because $m \geq 4$ is necessary to construct 
onion-like networks through intermediations 
without surface growth \cite{Hayashi18a}. 
Thus, it is the constraint for linking or forming a topological 
structure 
that attached nodes of one side of pair 
are not freely chosen uniformly at random, 
but chosen from only surface. 
In addition, according to another attachment in MED-kmin or MED-rand, 
destinations of $m/2$ links are chosen in the $(\mu + 1)$-th 
neighbors from the pairs on surface. 
If there is no $(\mu + 1)$-th neighbor, 
then the furthest node from its pair is chosen for a link. 
Note that links between a new node and the above neighbors make 
bridges whose lengths are more than unit length (one hop between 
contact nodes) of square lattice.
Such constructed network is not a planar graph, 
however the limitation of planarity is hardly
required in transportation or communication systems.

\section{Emergence of onion-like networks}
We use two measures of robustness index $R$ and assortativity $r$ 
for investigating the emergence of onion-like networks. 
The robustness index $1/N \leq R \leq 0.5$ \cite{Schneider11} 
is defined as $R \stackrel{\rm def}{=} \sum_{q=1/N}^{1} S(q) / N$, 
where $S(q)$ denotes the number of nodes included in the giant 
component (as the largest connected cluster) 
after removing $q N$ nodes, $q$ is a fraction of removed nodes. 
In this paper, we consider malicious attacks with 
recalculation of the largest degree node. 
For the measure of degree-degree correlations, 
as the Pearson correlation coefficient for degrees \cite{Newman03}, 
the assortativity $-1 \leq r \leq 1$ is defined as 
\[
  r \stackrel{\rm def}{=} \frac{4 M \sum_{e} (k_{e} k'_{e}) 
  - \left[ \sum_{e} (k_{e} + k'_{e}) \right]^{2}}{
  2 M \sum_{e} (k^{2}_{e} + k'^{2}_{e}) 
  - \left[ \sum_{e} (k_{e} + k'_{e}) \right]^{2}},
\]
where $k_{e}$ and $k'_{e}$ denote degrees at end-nodes 
of link $e$, $M$ is the total number of links. 
When both $R$ and $r$ are large, the network is onion-like.
Although there is no exact criterion to be onion-like, 
we investigate it in numerical comparison. 
The following results are averaged over $100$ realizations 
of generated networks from the initial $K_{m}$ at the origin 
$(0, 0)$ of square lattice. 

\begin{figure}[htb]
\begin{tabular}{cc}
 Robustness \hspace{3cm} Correlation
 \\
 \begin{minipage}{0.5\hsize}
   \includegraphics[width=0.7\linewidth,angle=-90]{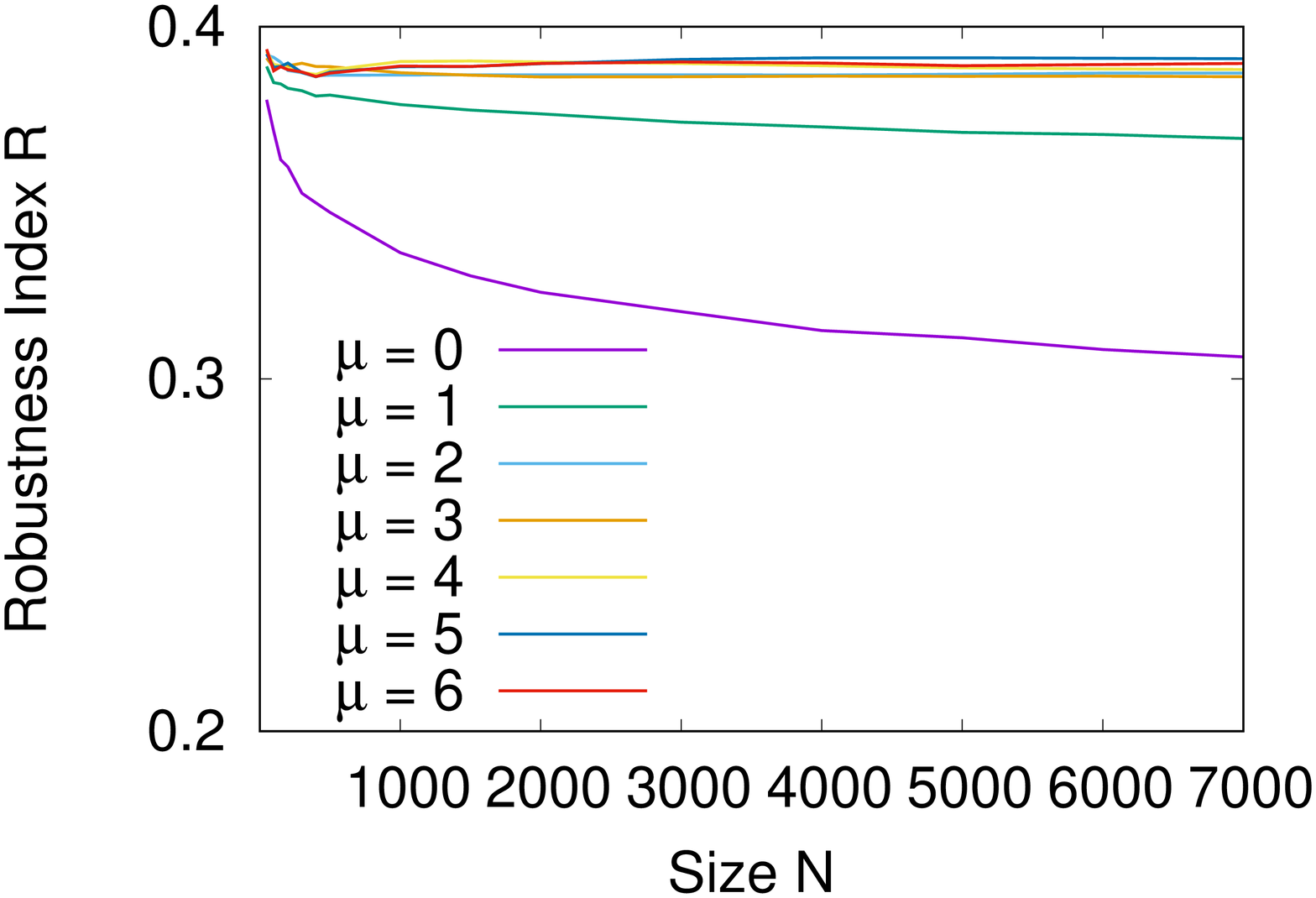}
 \end{minipage}
 \begin{minipage}{0.5\hsize}
   \includegraphics[width=0.7\linewidth,angle=-90]{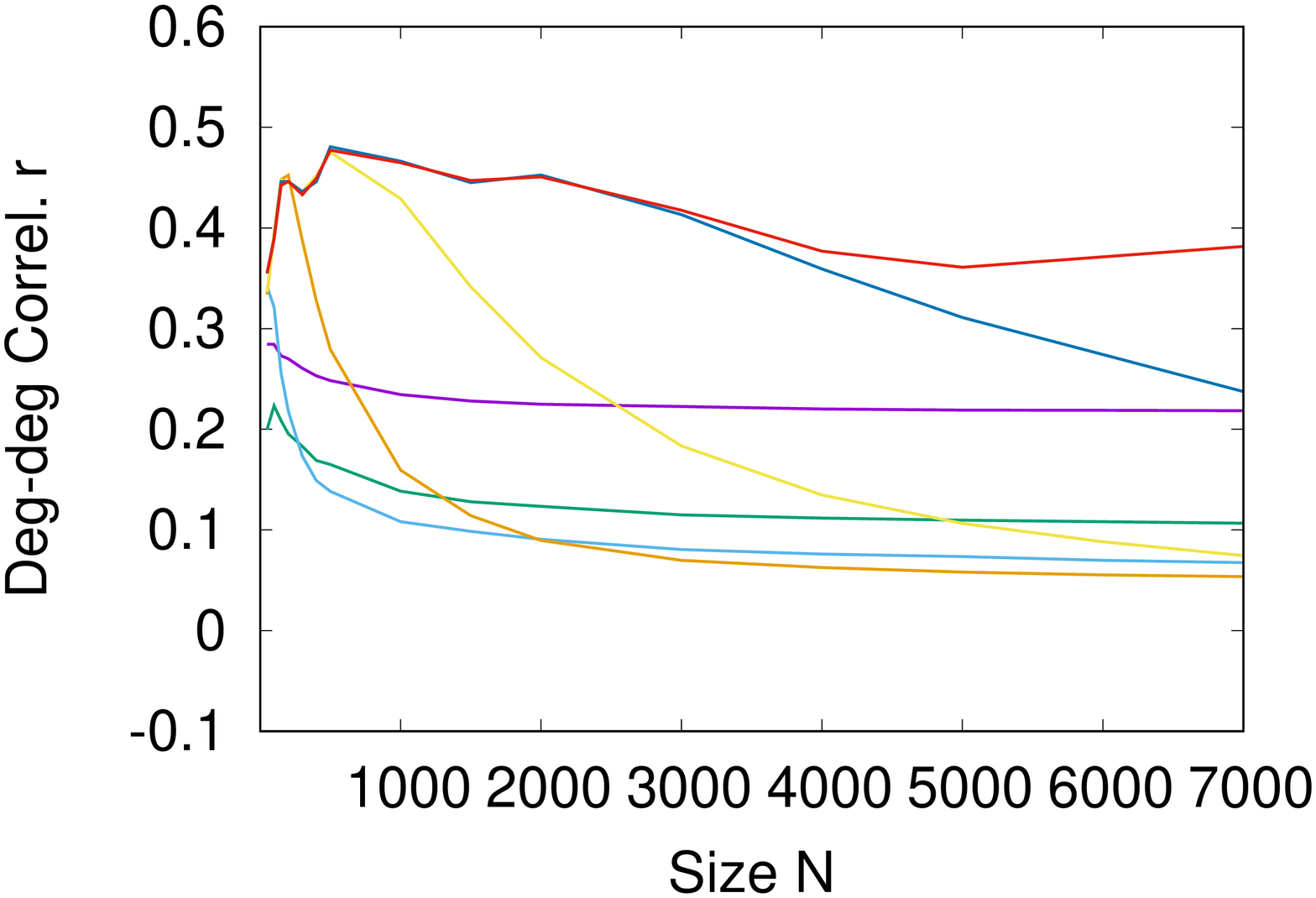}
 \end{minipage}
 \\
 \begin{minipage}{0.5\hsize}
   \includegraphics[width=0.7\linewidth,angle=-90]{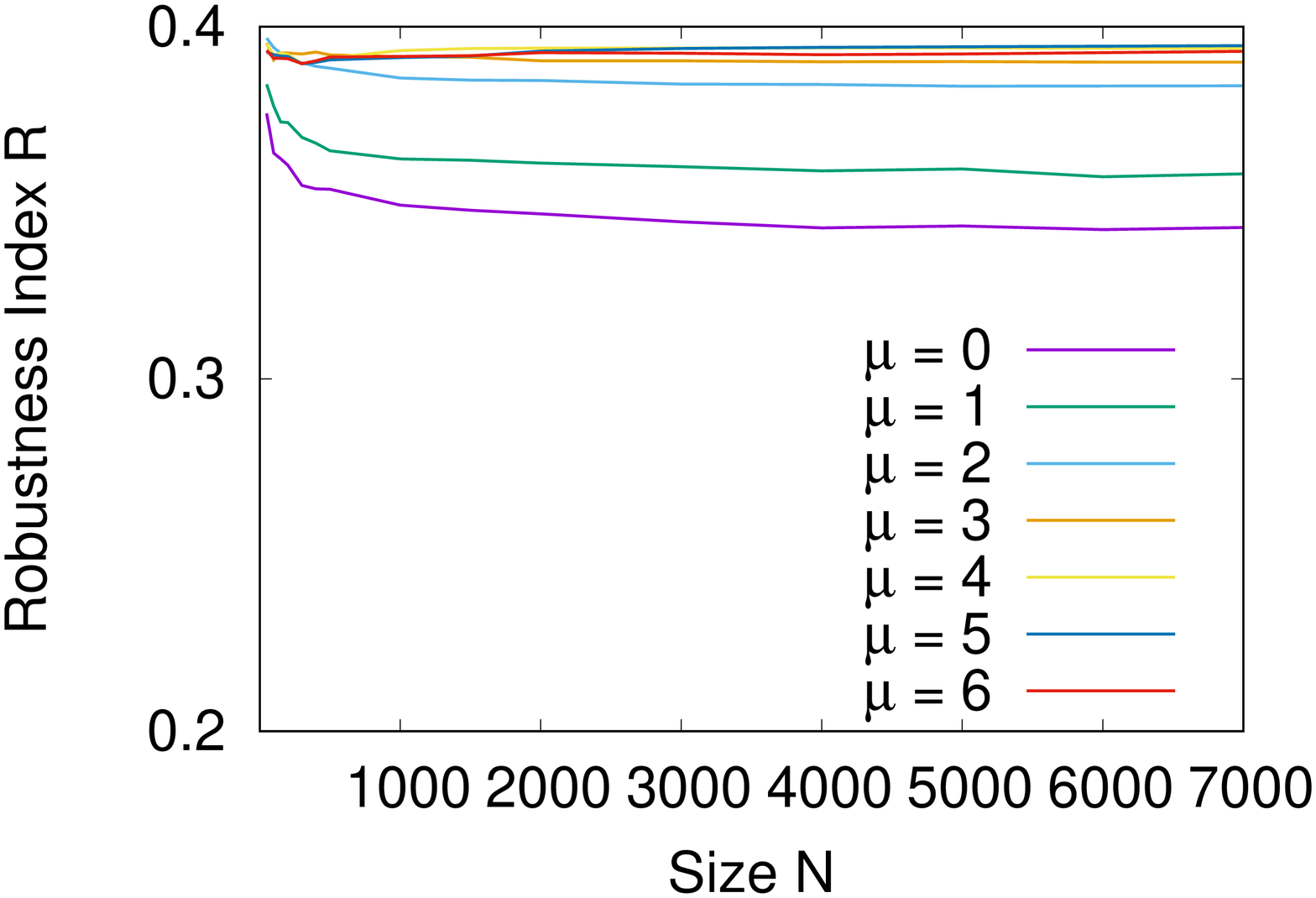}
 \end{minipage}
 \begin{minipage}{0.5\hsize}
   \includegraphics[width=0.7\linewidth,angle=-90]{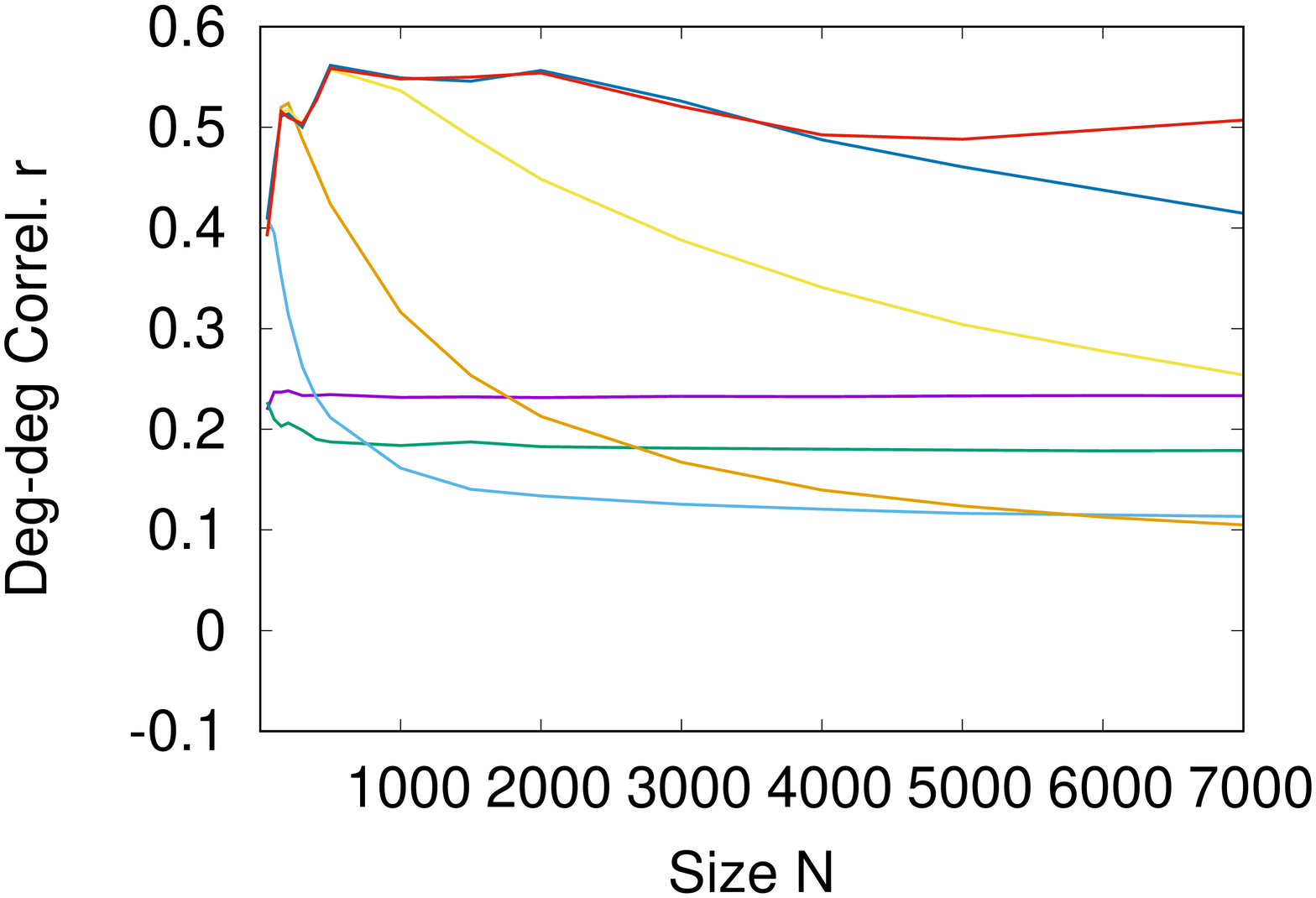}
 \end{minipage}
 \\
 \begin{minipage}{0.5\hsize}
   \includegraphics[width=0.7\linewidth,angle=-90]{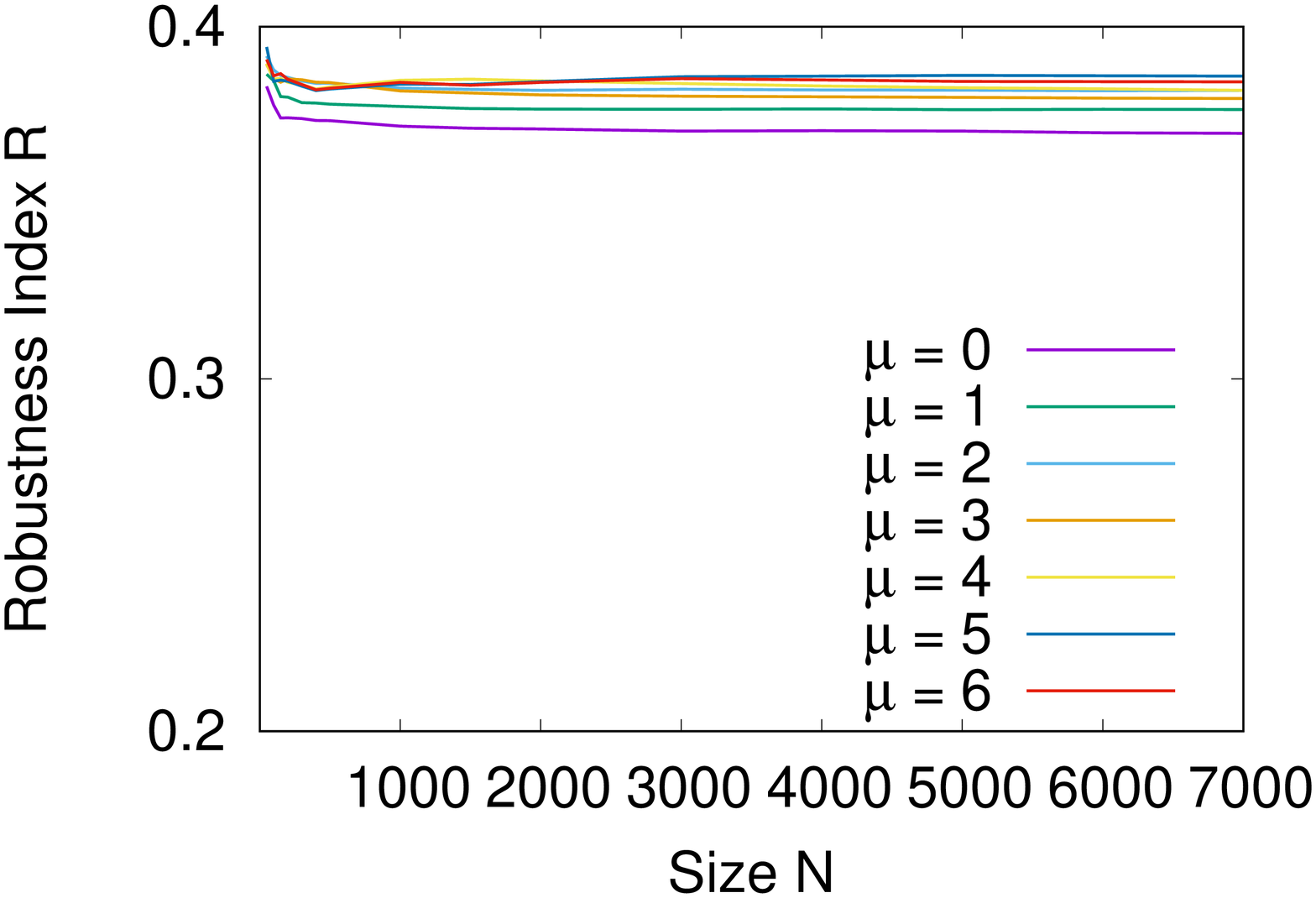}
 \end{minipage}
 \begin{minipage}{0.5\hsize}
   \includegraphics[width=0.7\linewidth,angle=-90]{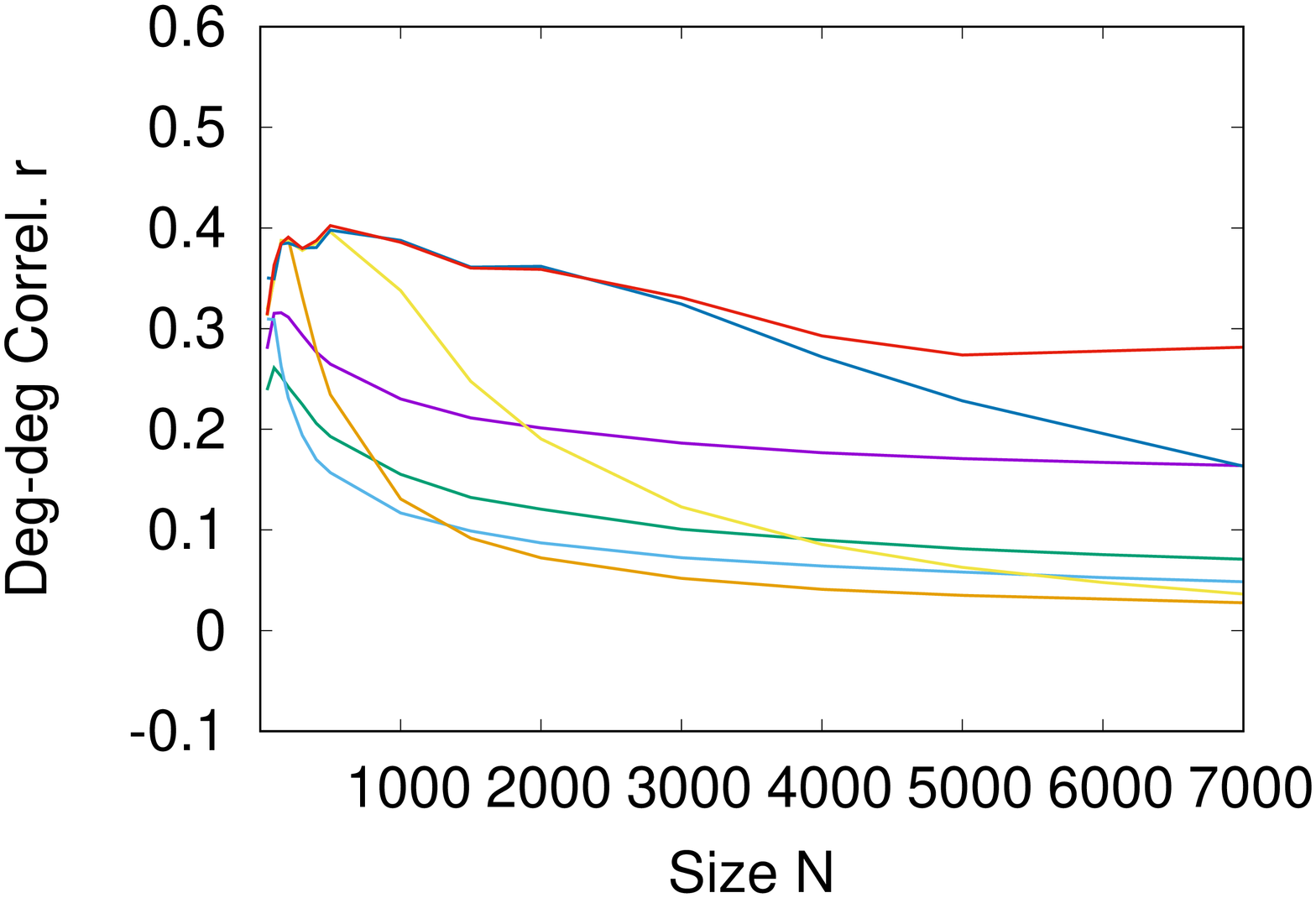}
 \end{minipage}
\end{tabular}
\caption{Results for the method of MED-kmin with surface constraint 
by DLA(Top), IP(Middle), and Eden(Bottom) models.}
\label{fig_kmin_R_r}
\end{figure}

Figure \ref{fig_kmin_R_r} shows the results for 
the growing method of MED-kmin. 
The robustness is almost constantly strong around $R > 0.37$ 
except the cases of $\mu = 0, 1$ (purple and green lines). 
Similar robustness is obtained in the three models of surface growth. 
For the degree-degree correlations, 
only the cases of $\mu = 6$ (red line) by DLA model 
and $\mu = 5, 6$ (blue and red lines) by IP model 
have $r > 0.36$. 
The case of $\mu = 6$ (red line) by Eden model 
have slightly small $R \approx 0.3$. 
In other cases, $r$ is small around $0.1 \sim 0.2$.
Here, these $R \approx 0.37$ and $r \approx 0.36$ 
are the same level in the original method of 
MED-kmin without surface constraint \cite{Hayashi18a,Hayashi18b}. 

\begin{figure}[htb]
\begin{tabular}{cc}
 Robustness \hspace{3cm} Correlation
 \\
 \begin{minipage}{0.5\hsize}
   \includegraphics[width=0.7\linewidth,angle=-90]{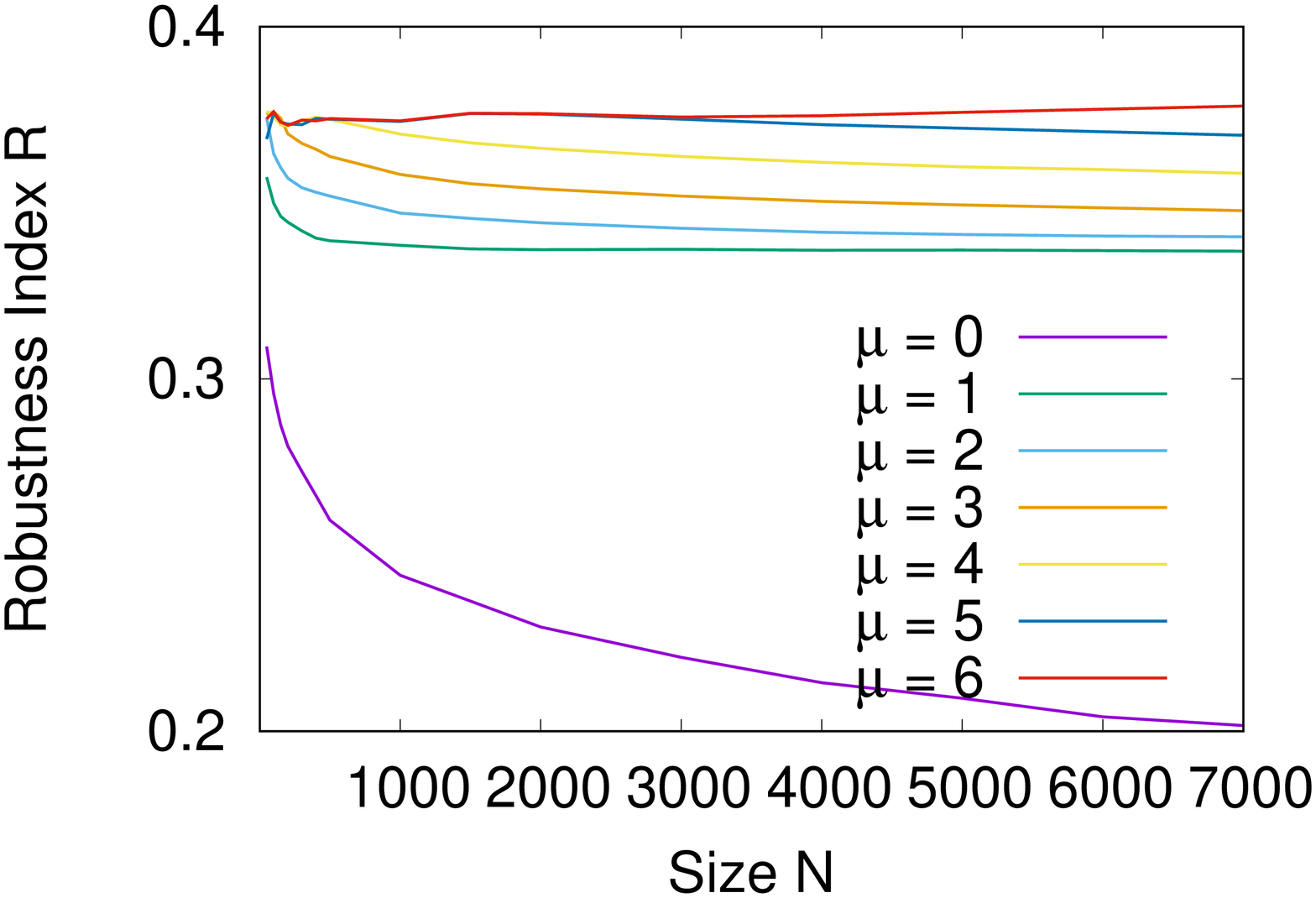}
 \end{minipage}
 \begin{minipage}{0.5\hsize}
   \includegraphics[width=0.7\linewidth,angle=-90]{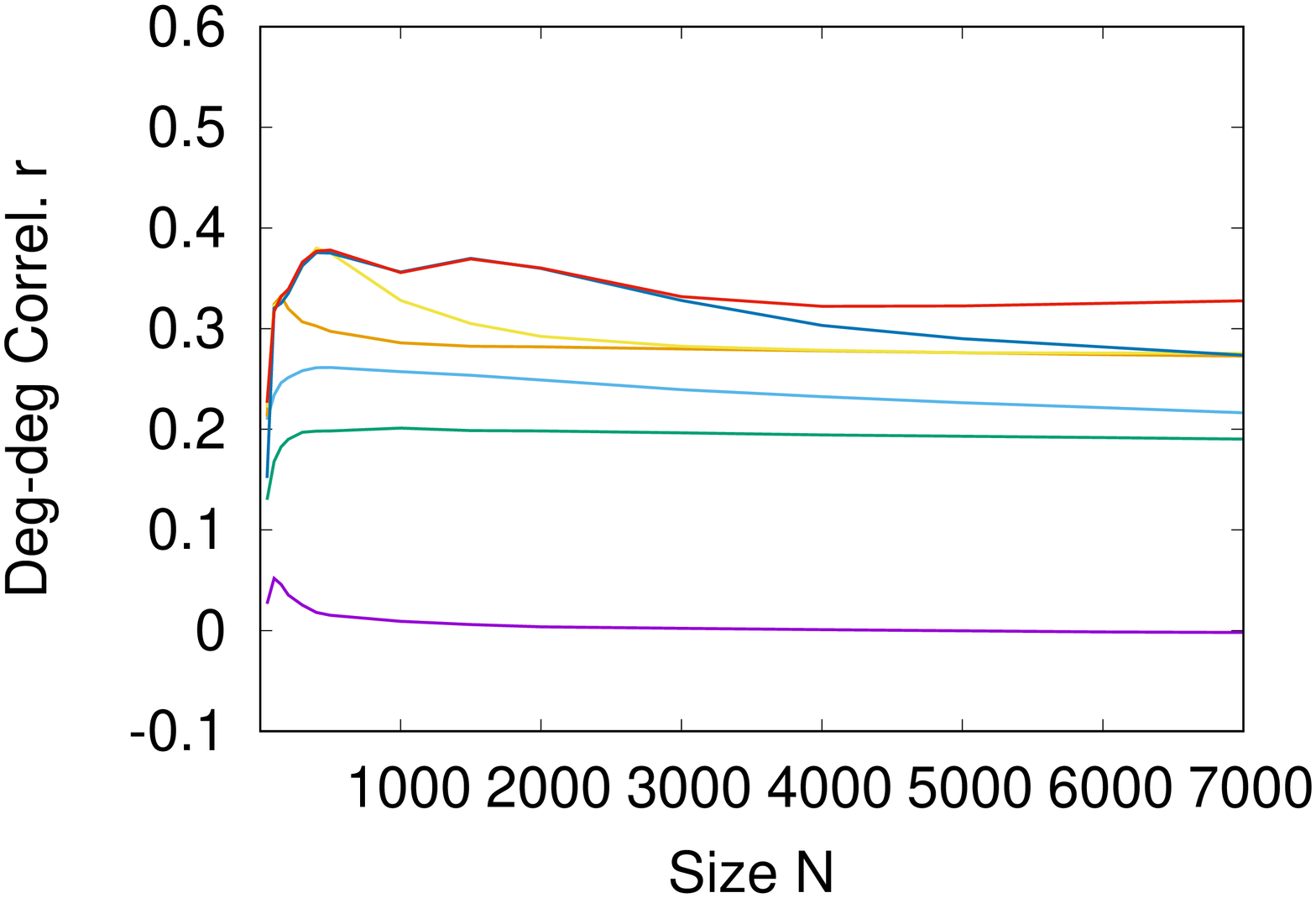}
 \end{minipage}
 \\
 \begin{minipage}{0.5\hsize}
   \includegraphics[width=0.7\linewidth,angle=-90]{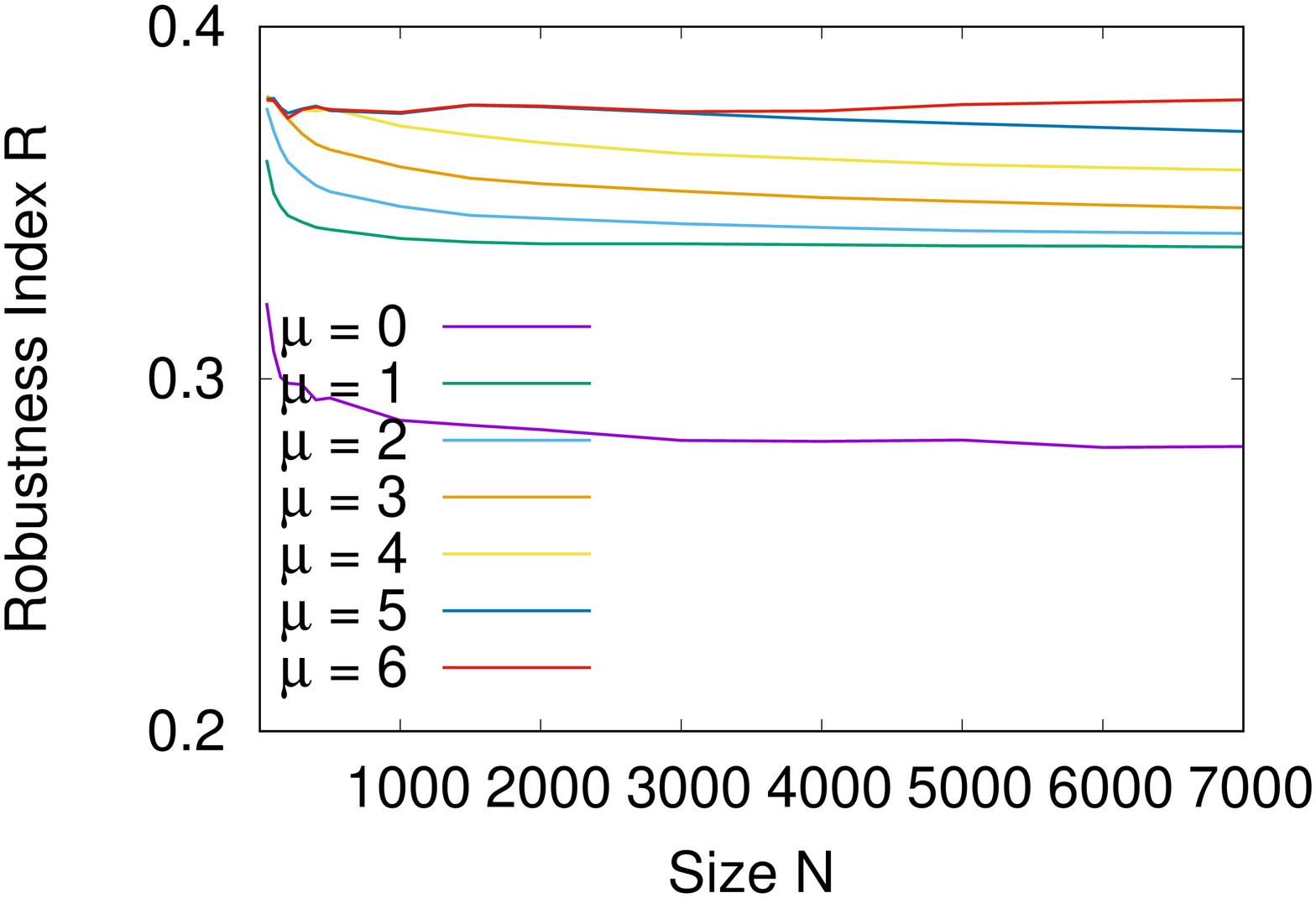}
 \end{minipage}
 \begin{minipage}{0.5\hsize}
   \includegraphics[width=0.7\linewidth,angle=-90]{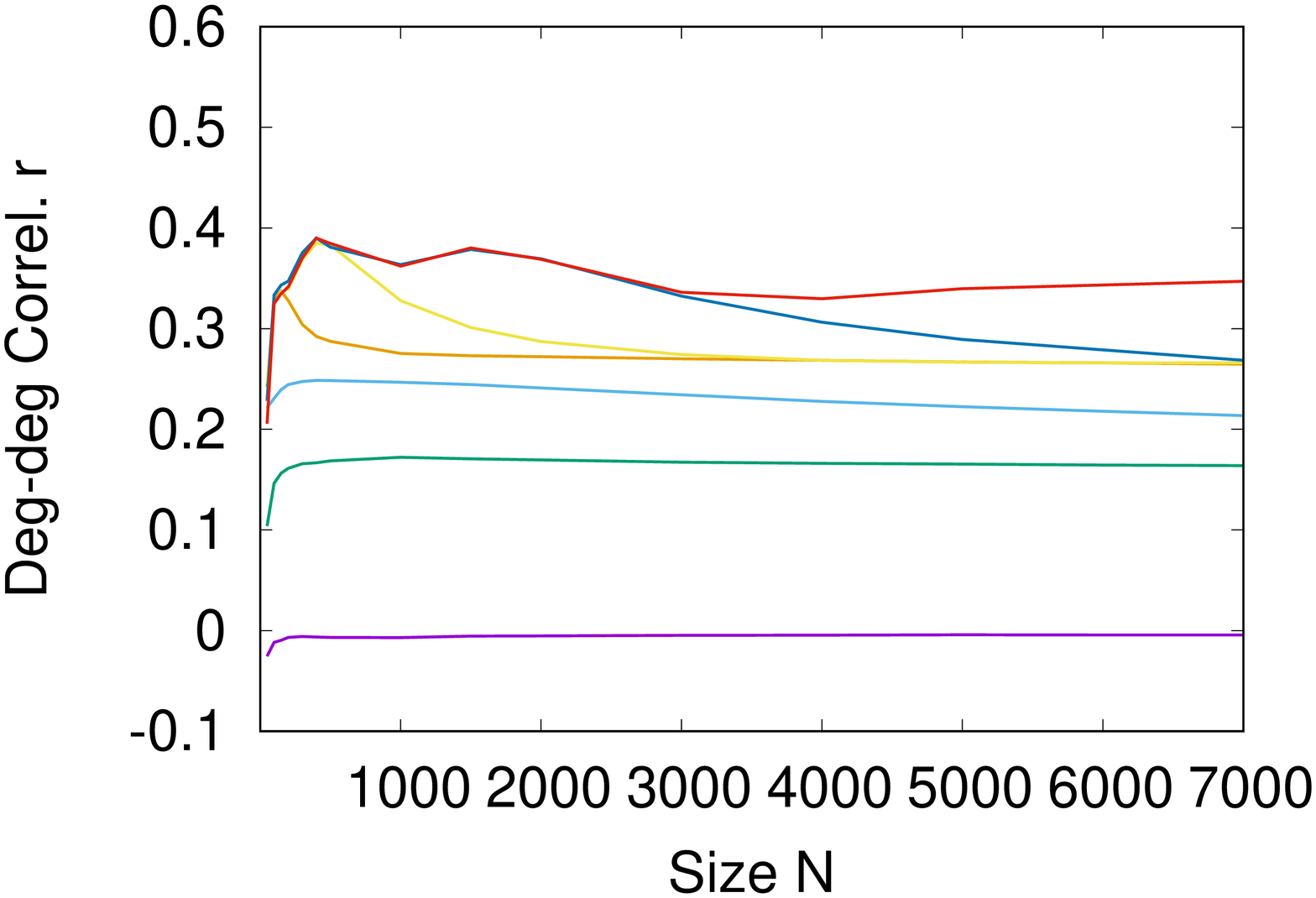}
 \end{minipage}
 \\
 \begin{minipage}{0.5\hsize}
   \includegraphics[width=0.7\linewidth,angle=-90]{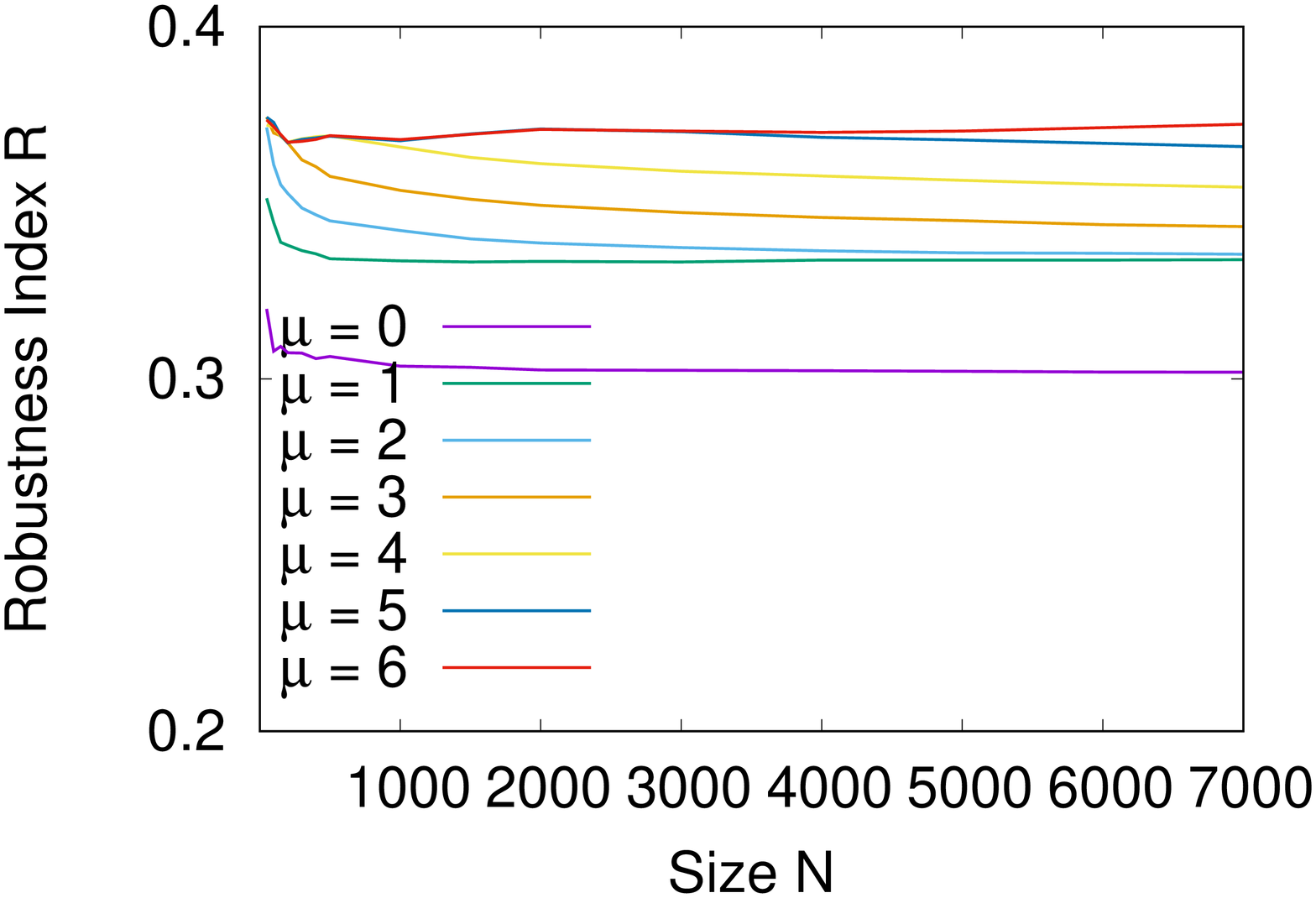}
 \end{minipage}
 \begin{minipage}{0.5\hsize}
   \includegraphics[width=0.7\linewidth,angle=-90]{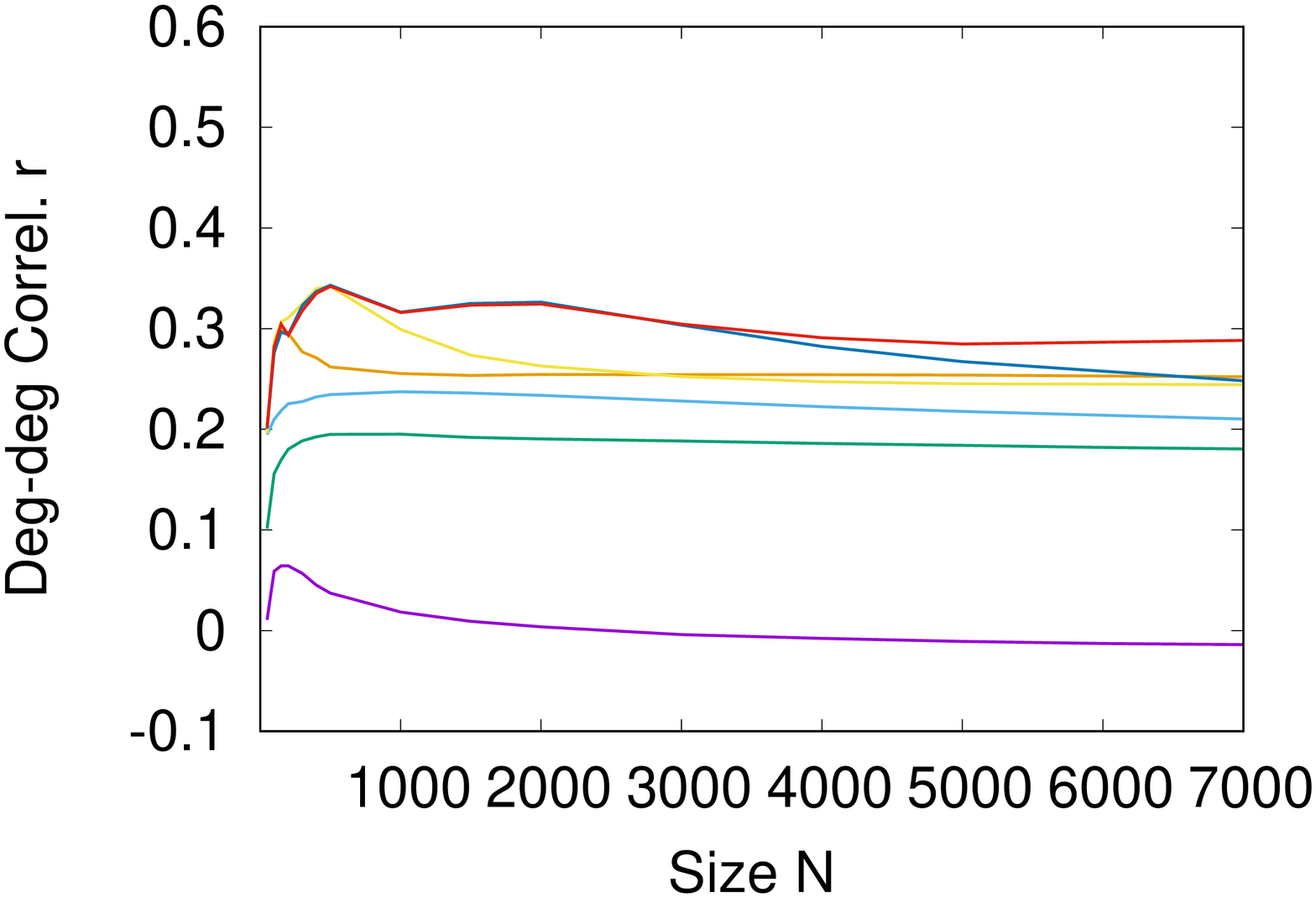}
 \end{minipage}
\end{tabular}
\caption{Results for the method of MED-rand with surface constraint 
by DLA(Top), IP(Middle), and Eden(Bottom) models.}
\label{fig_rand_R_r}
\end{figure}

Figure \ref{fig_rand_R_r} shows the results for 
the growing method of MED-rand. 
The robustness is also strong around $R > 0.34$ 
except the cases of $\mu = 0, 1, 2$ (purple, green, light blue lines). 
The robustness becomes larger as $\mu$ is larger in all 
three models of surface growth, 
the ordering of robustness by slight differences 
is Eden, IP, and DLA models from bottom to top. 
For the degree-degree correlations, 
the cases of $\mu = 4, 5, 6$ (yellow, blue, and red lines) 
have $r > 0.25$. 
The case of $\mu = 0$ (purple line) are almost no correlation of $r \approx 0$.
Here, these $R \approx 0.34$ and $r \approx 0.25$ 
are the same level in the original method of 
MED-rand without surface constraint \cite{Hayashi18a,Hayashi18b}. 
Thus, for the methods of MED-kmin and MED-rand of $\mu = 6$ 
intermadiations, onion-like networks emerge with large $R$ and $r$ 
in the constraint of surface growth. 
Even if the network after growing until each size $N$
is rewired by Wu-Holme method \cite{Wu11}, 
the robustness is similar around $R \approx 0.38 \sim 0.39$. 
Note that the copying networks with surface constraint 
have $R \approx 0.27$ and $r \approx 0.3$ \cite{Hayashi16}, 
however the robustness is weaker as $R < 0.2$ 
in early stage of $N < 1000$.

\begin{figure}[htb]
\begin{tabular}{cc}
 MED-kmin \hspace{3cm} MED-rand
 \\
 \begin{minipage}{0.5\hsize}
   \includegraphics[width=0.7\linewidth,angle=-90]{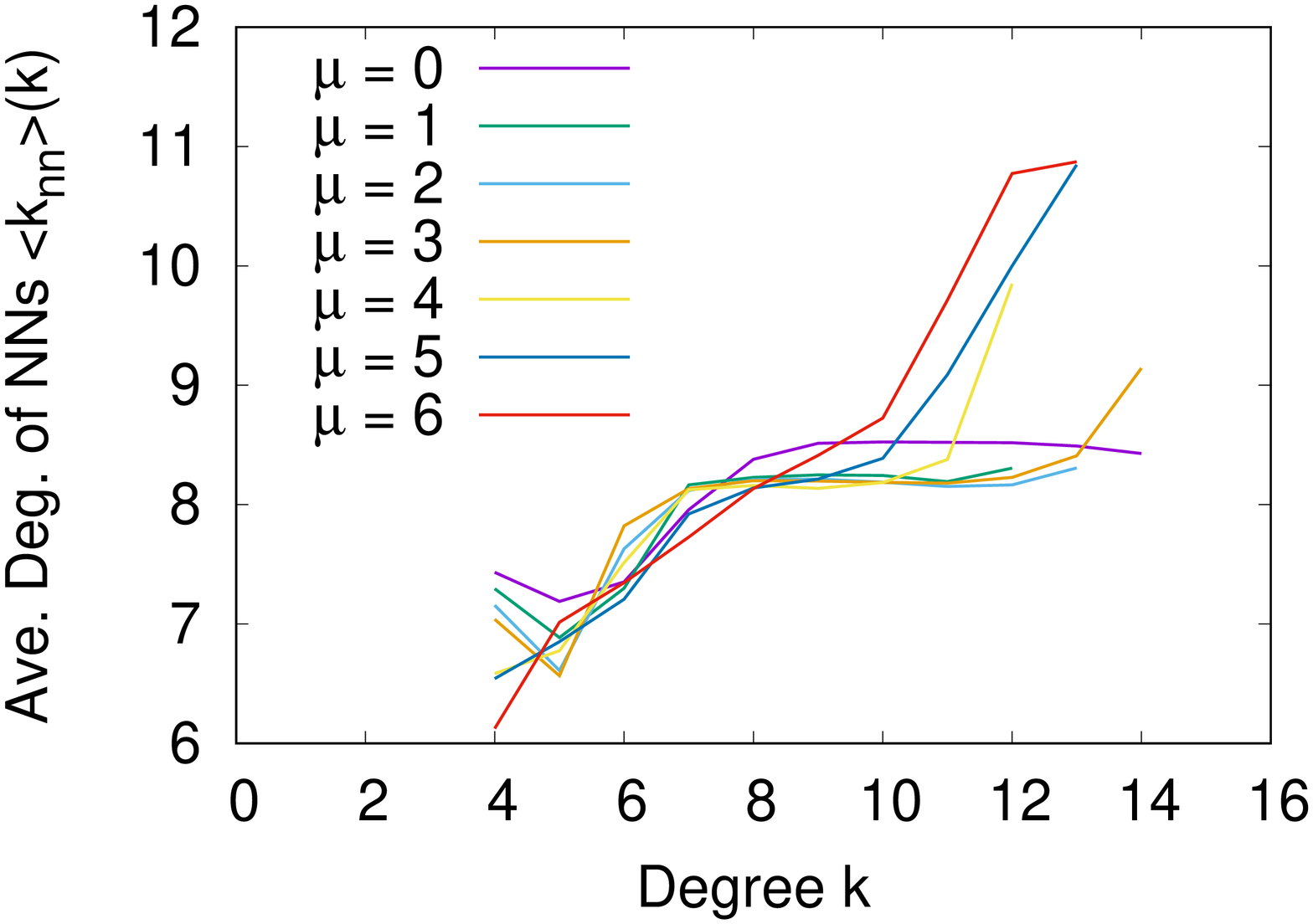}
 \end{minipage}
 \begin{minipage}{0.5\hsize}
   \includegraphics[width=0.7\linewidth,angle=-90]{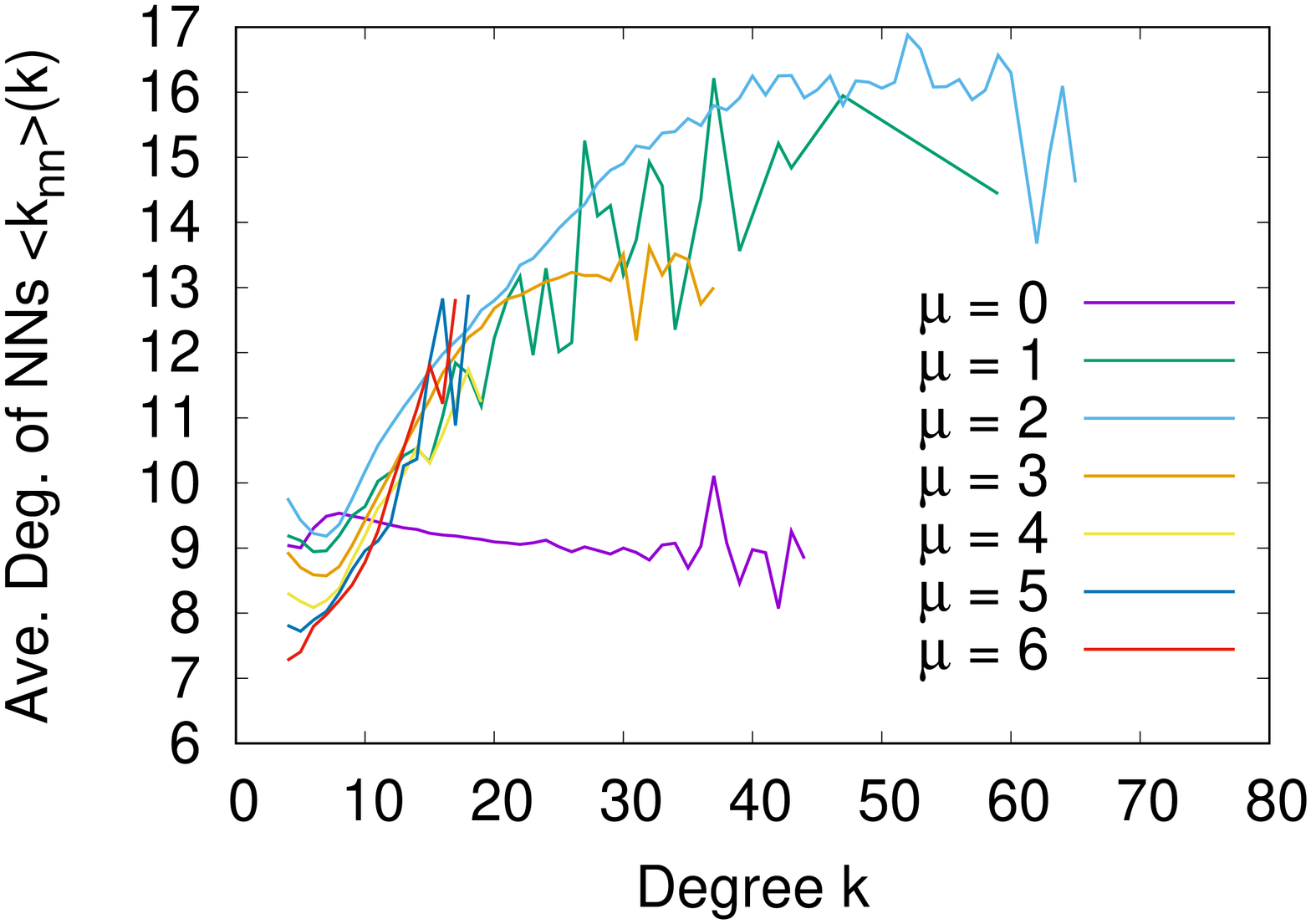}
 \end{minipage}
 \\
 \begin{minipage}{0.5\hsize}
   \includegraphics[width=0.7\linewidth,angle=-90]{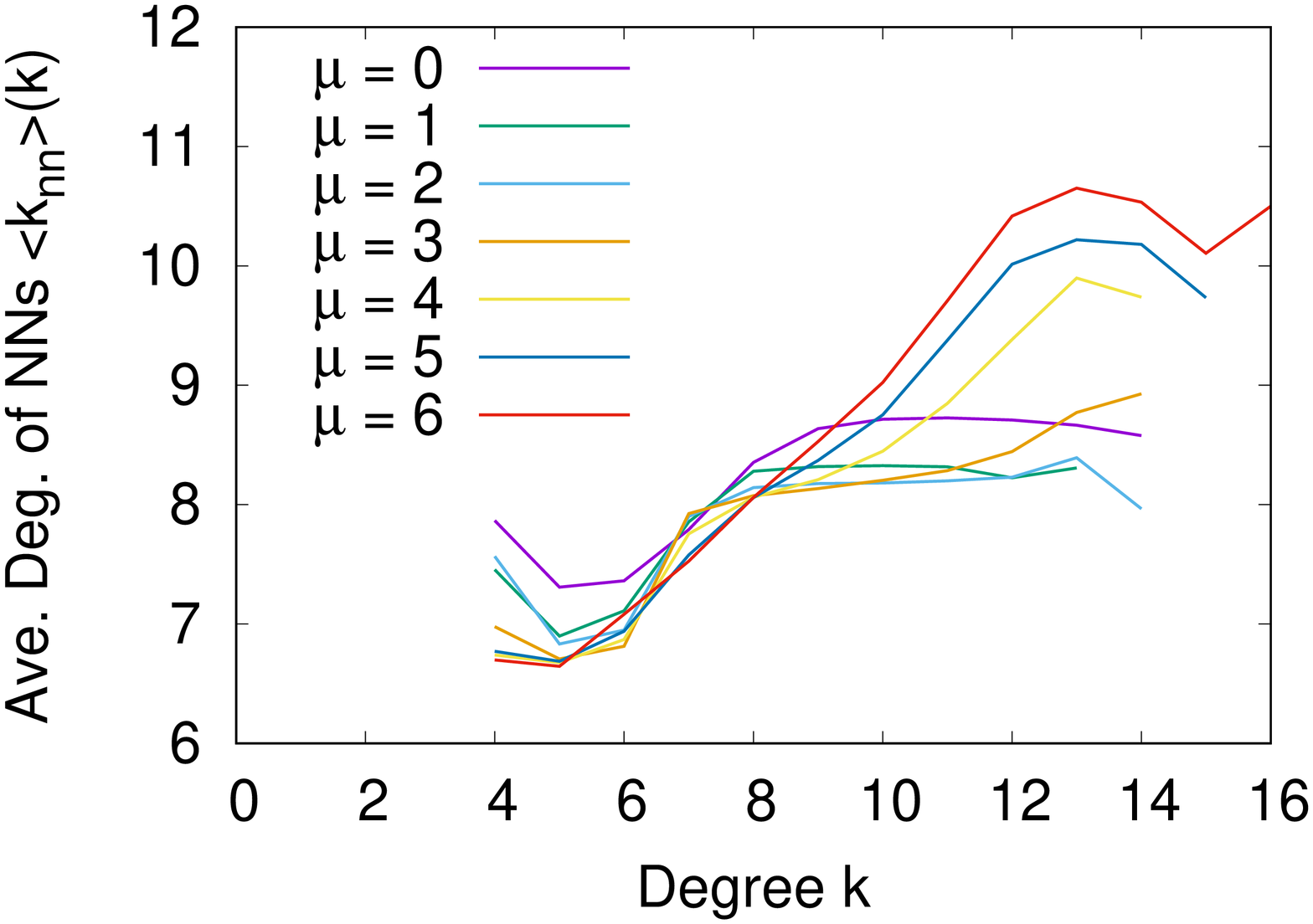}
 \end{minipage}
 \begin{minipage}{0.5\hsize}
   \includegraphics[width=0.7\linewidth,angle=-90]{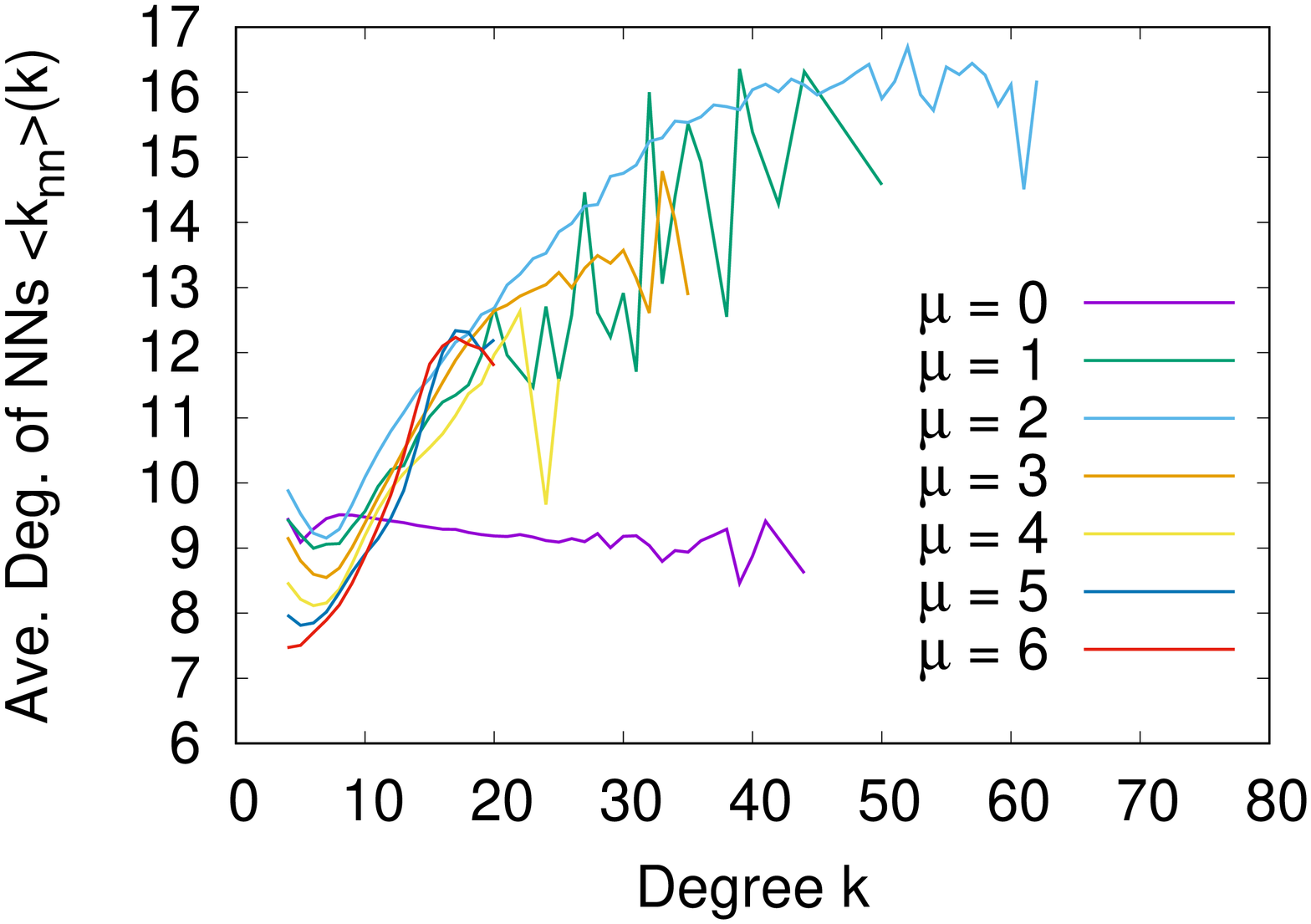}
 \end{minipage}
 \\
 \begin{minipage}{0.5\hsize}
   \includegraphics[width=0.7\linewidth,angle=-90]{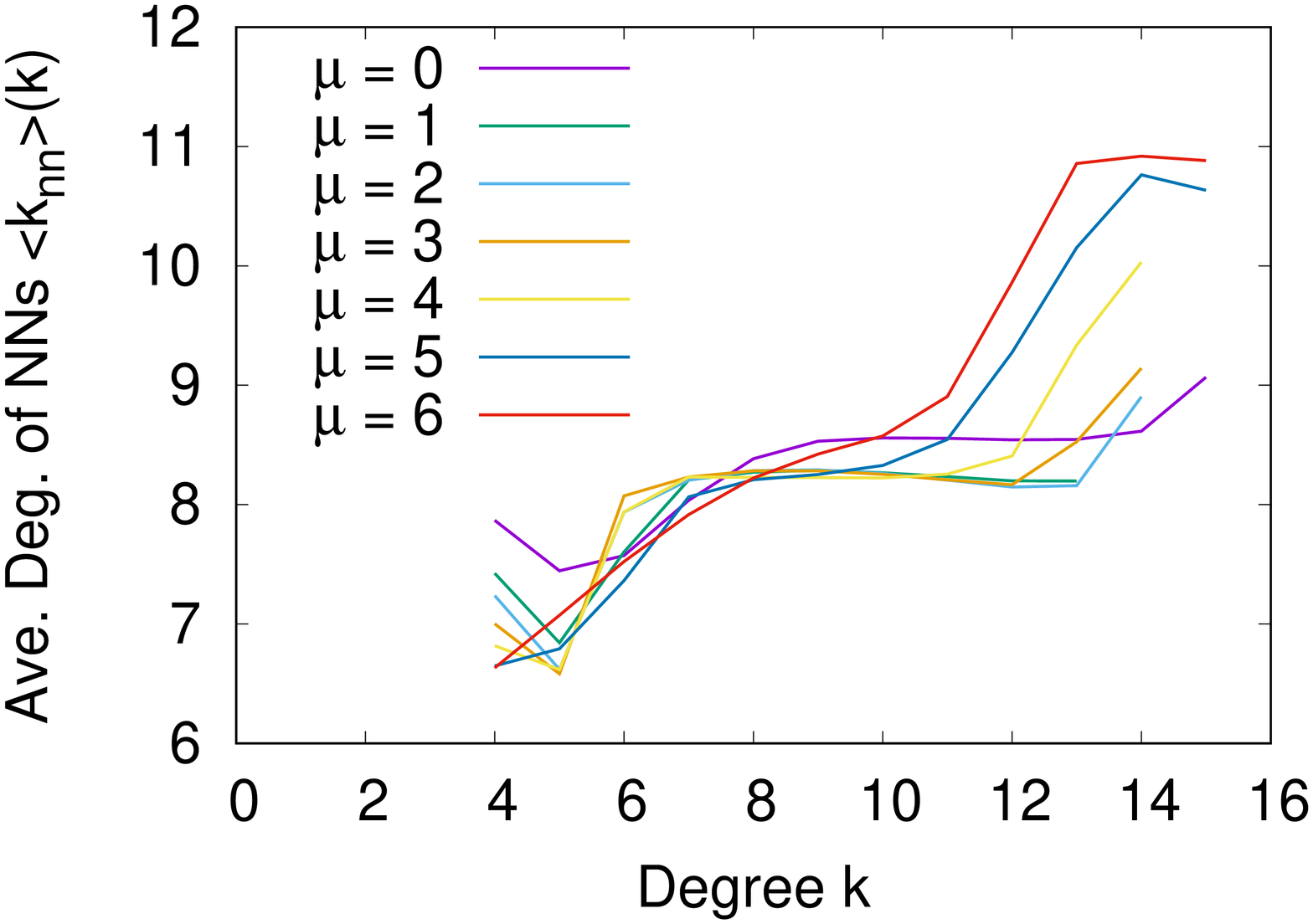}
 \end{minipage}
 \begin{minipage}{0.5\hsize}
   \includegraphics[width=0.7\linewidth,angle=-90]{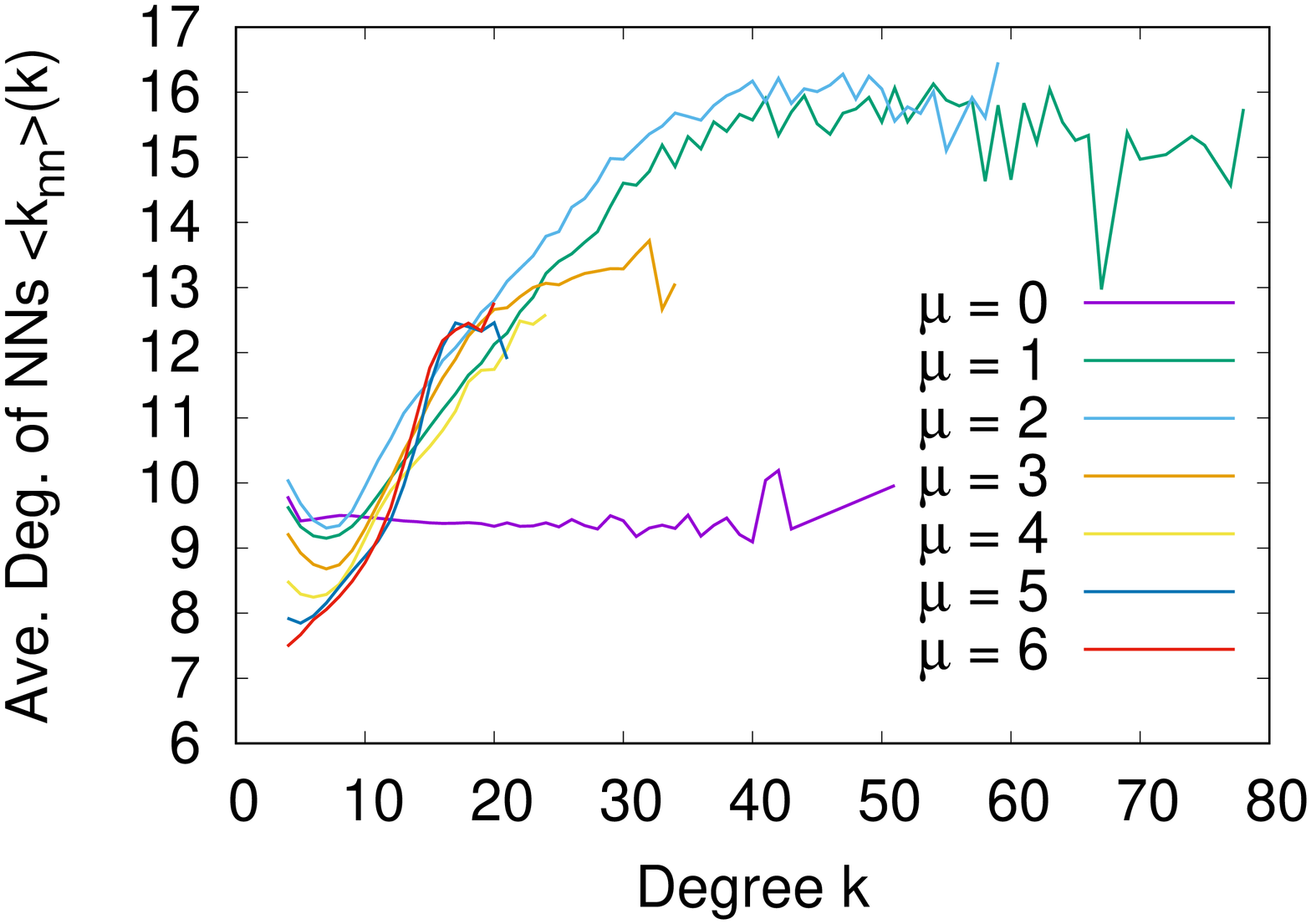}
 \end{minipage}
\end{tabular}
\caption{ADNN in the networks at $N=7000$ 
with surface constraint by DLA(Top), IP(Middle), and Eden(Bottom) models.}
\label{fig_k_nn}
\end{figure}

Moreover, we consider the reason why onion-like networks emerge 
in the methods. 
Figure \ref{fig_k_nn} shows the 
average degree of the nearest neighbors (ADNN) 
$\langle k_{nn} \rangle(k)$ as a function of $k$.
As the slope $\Delta \langle k_{nn} \rangle / \Delta k$ close to one, 
similar degree nodes tend to connect, then 
the degree-degree correlations are enhanced. 
In contrast, as negative one, the correlations are inhibited. 
In the method of MED-kmin of $\mu = 6$ 
(red line in the left of Fig. \ref{fig_k_nn}), 
positive slopes are dominant, 
while there are plateaus which weaken the correlations 
in other cases of $\mu = 0, 1, \ldots, 5$ (purple $\sim$ blue lines).
The plateaus at $\langle k_{nn} \rangle(k) \approx 8$ which is 
corresponded to the maximum contacts $8$ in random attachment 
are longer as $\mu$ is smaller. 
Similarly, in the method of MED-rand of $\mu = 6$ 
(red line in the right of Fig. \ref{fig_k_nn}), 
positive slopes are dominant. 
Mixed positive and negative slopes exist in other cases of 
$\mu = 0, 1, \ldots, 5$. In particular, the slope is almost flat 
in the case of $\mu = 0$ (purple line), and the flat part occupies 
about half of the line 
in the cases of $\mu = 1, 2$ (green and light blue lines). 
These length of plateau or frequency of negative slops 
correspond to the ordering of suppressed $r$ values 
in Figs. \ref{fig_kmin_R_r},\ref{fig_rand_R_r}. 
We remark that the maximum degree in the method of MED-rand 
is larger than that in the method of MED-kmin.

\begin{figure}[htb]
\centering
\begin{tabular}{c}
 \begin{minipage}{0.5\hsize}
 \centering
 MED-kmin \hspace{2mm} MED-rand
 \vspace{1.5mm}

   \includegraphics[width=0.9\linewidth]{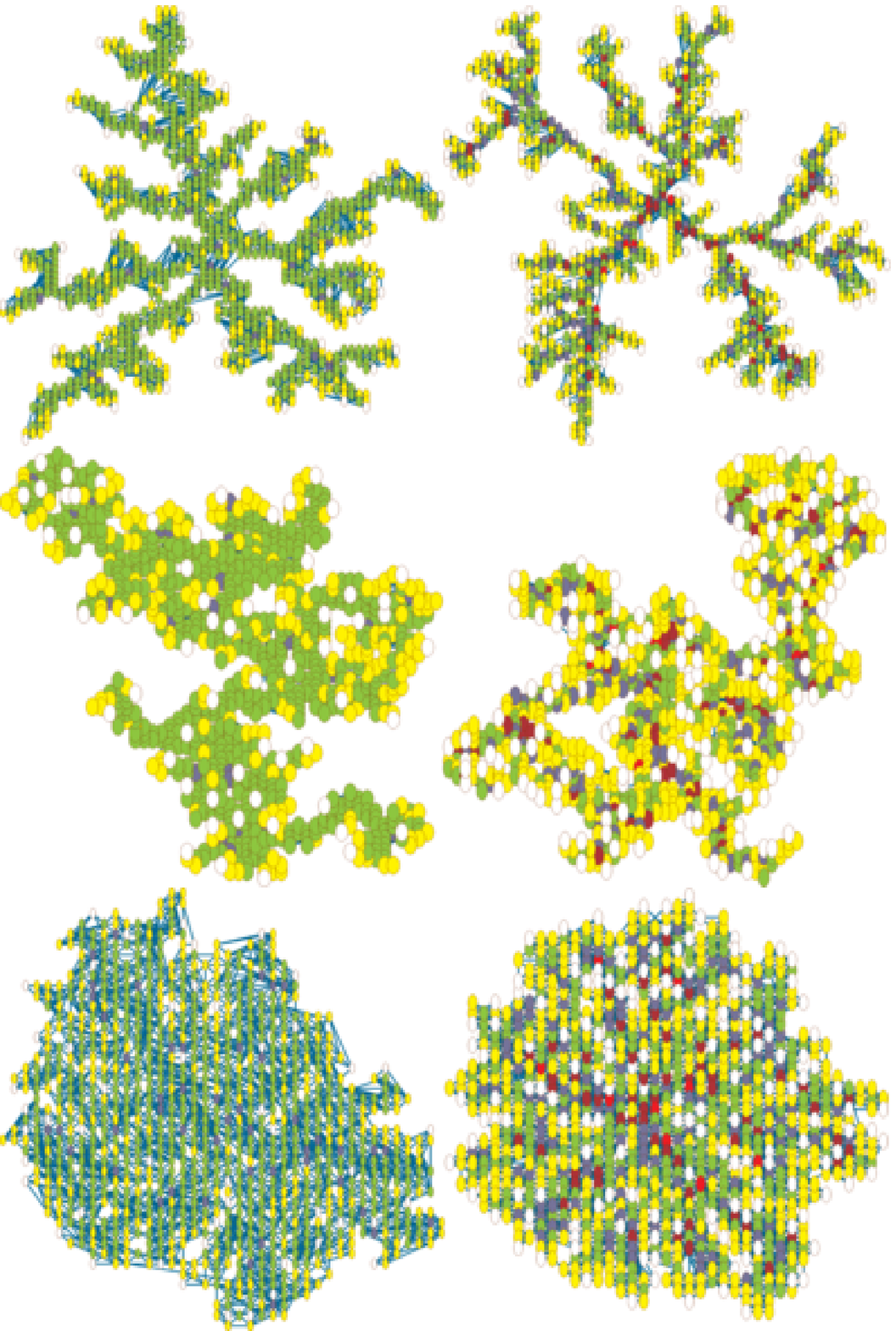}

 (a) $\mu = 0$
 \end{minipage}
 
 \begin{minipage}{0.5\hsize}
 \centering
 MED-kmin \hspace{2mm} MED-rand
 \vspace{1.5mm}

   \includegraphics[width=0.9\linewidth]{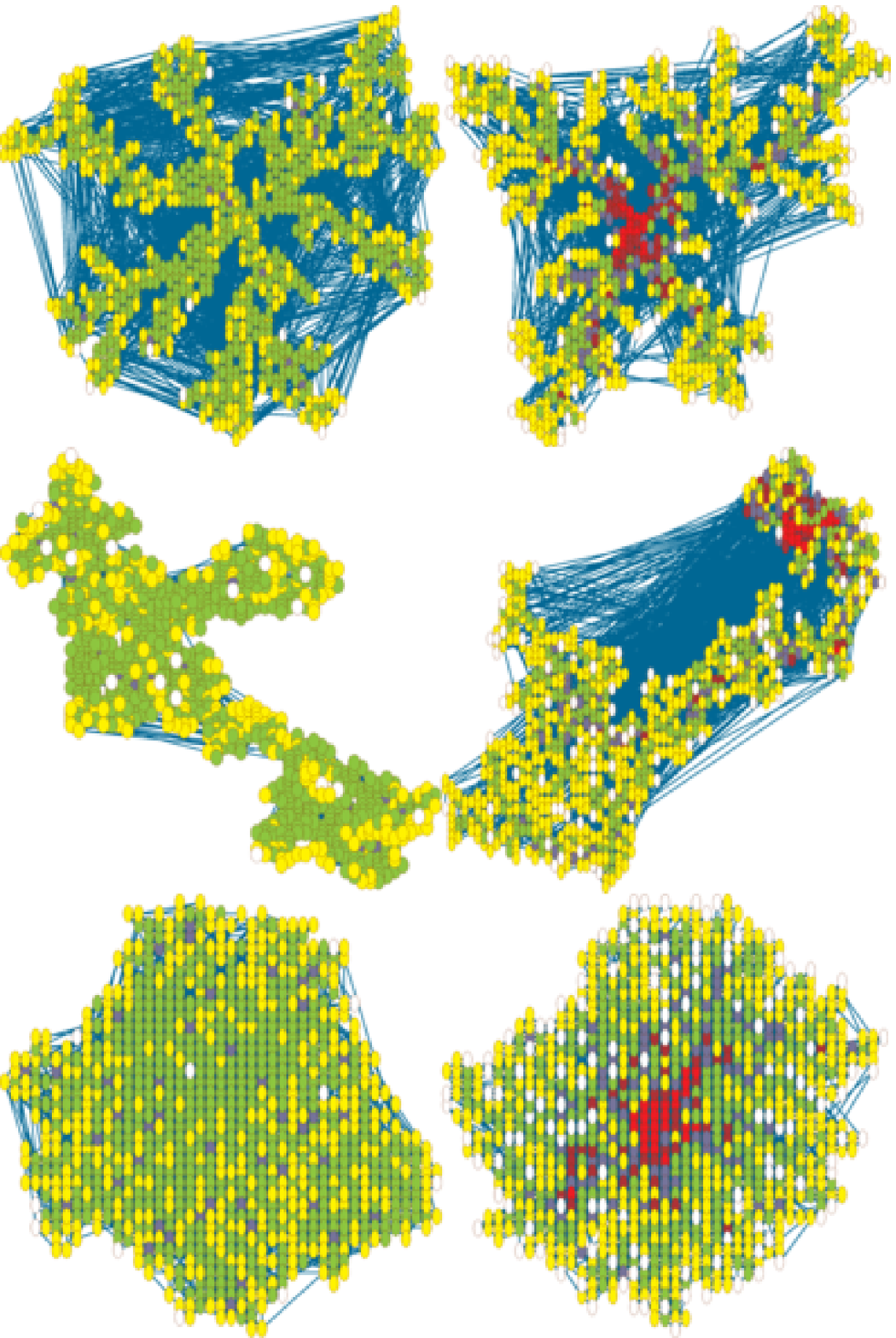}

 (b) $\mu = 1$
 \end{minipage}
\end{tabular}

\vspace{3mm}
\begin{tabular}{c}
 \begin{minipage}{0.5\hsize}
 \centering
 MED-kmin \hspace{2mm} MED-rand
 \vspace{1.5mm}

   \includegraphics[width=0.9\linewidth]{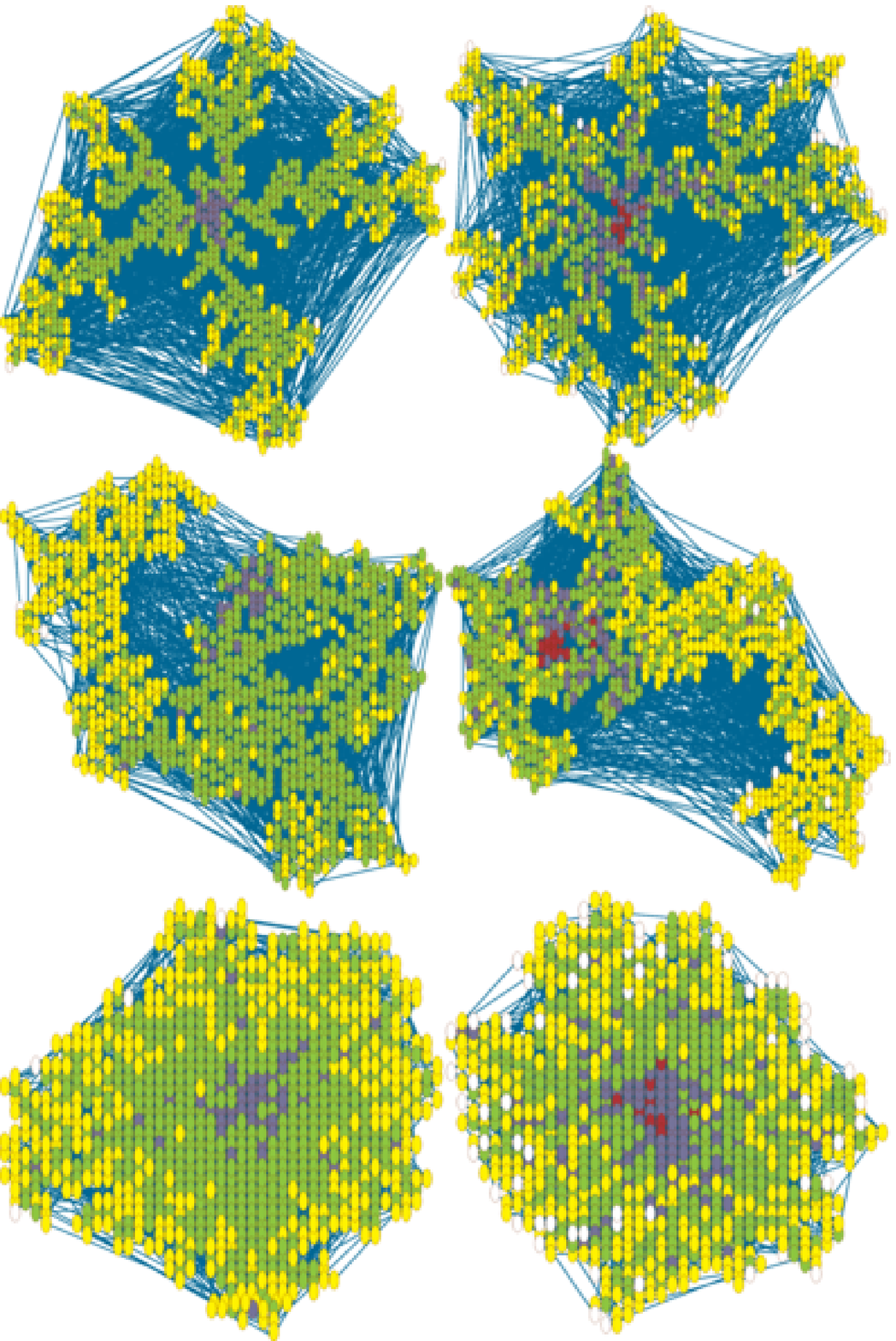}

 (c) $\mu = 5$
 \end{minipage}
 
 \begin{minipage}{0.5\hsize}
 \centering
 MED-kmin \hspace{2mm} MED-rand
 \vspace{1.5mm}

   \includegraphics[width=0.9\linewidth]{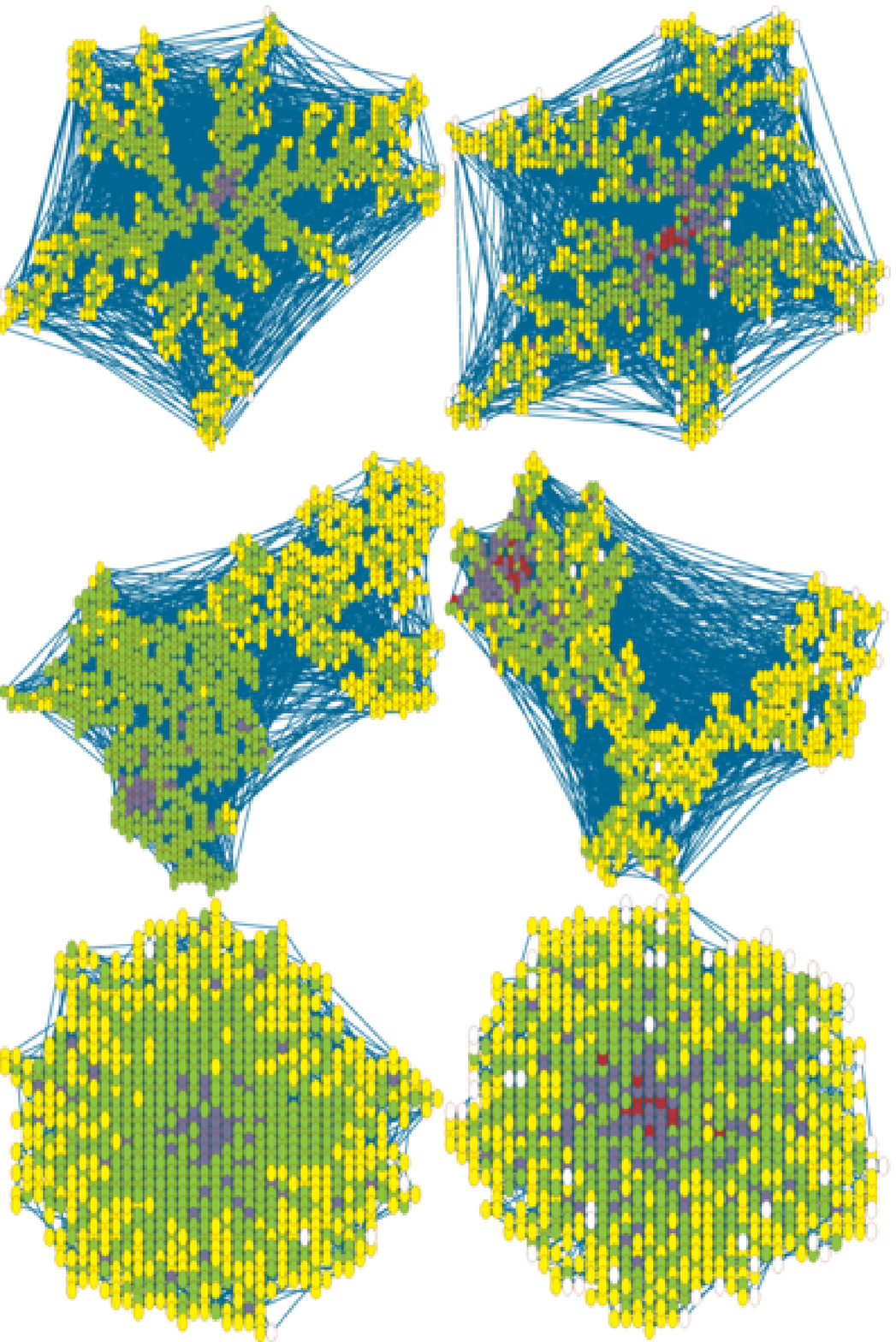}

 (d) $\mu = 6$
 \end{minipage}
\end{tabular}
\caption{Existence of the inner core at $N = 1000$.}
\label{fig_vis}
\end{figure}

The strong correlations shown by $\langle k_{nn} \rangle$ in 
Fig. \ref{fig_k_nn} correspond to the existence of inner core of 
large degree nodes in the network with surface constraint by 
DLA(Top), IP(Middle), and Eden(Bottom) models.
In Fig. \ref{fig_vis}, 
colors of nodes are assigned for each range of degrees; 
$4$: white, $5, 6, 7$: yellow, $8, 9, 10$: green, 
$11, 12, 13$: gray, $14, \ldots, 19$: brown, and $20$ over: red. 
Blue line denotes a link. 
In the method of MED-kmin, 
there exists no core of large degree nodes for intermediations of 
$\mu = 0, 1$ in Fig. \ref{fig_vis}(a)(b), 
however it emerges as a gray area for $\mu = 5, 6$ in 
Fig. \ref{fig_vis}(c)(d). 
It is difficult to see 
any difference from the visualization, 
although $\langle k_{nn} \rangle$ in 
Fig. \ref{fig_k_nn} (blue and red lines) slightly differs 
for $\mu = 5$ and $\mu = 6$. 
Thus, moderately long links through intermediations 
are necessary in order to connect 
from a new node on surface to the core.
We should remark that 
older nodes are closer to the core, and get links by 
intermediation attachment, 
while newer nodes are closer to surface, and get links 
by random attachment, mainly.
The roles of  enhancing the correlations for large nodes 
and small nodes are exchanged from them 
in the original method \cite{Hayashi18a,Hayashi18b} 
without constraint of surface growth.

In the method of MED-rand, 
large degree (brown and red) nodes appear in the inner core 
as shown in Fig. \ref{fig_vis}(a)-(d). 
There are several common features for MED-kmin and MED-rand. 
In the case by IP model, 
the position of core (gray or brown) with concentrated links 
is not determined in advance but by chance 
as shown at the middle in Fig. \ref{fig_vis}(a)-(d).
In addition, 
the degree becomes medium (green) in the area out of the core, 
while it tend to be smaller (yellow) as closer to surface.
Outer (blue) links between surface nodes appear in $\mu \geq 1$
as shown from Fig. \ref{fig_vis}(b) to (d), 
while it is a reason of weak robustness in the case of $\mu=0$ 
that outer links does not appear in Fig. \ref{fig_vis}(a). 
Thus, outer links contribute to enhance 
the robustness of connectivity, 
moderately long links are also necessary in this reason.

\section{Conclusion}
We have shown that onion-like topological structure 
with positive degree-degree correlations 
and strong robustness of connectivity can emerge in a spatially 
embedded network by pair of random and intermediation attachments 
even with the constraint of surface growth.
In particular, it is found that 
moderately long links are necessary to enhance the correlations 
between large degree nodes at the inner core 
and to be onion-like.
These results will be useful for designing prospective 
future networks for transportation or communication systems.

\section*{Acknowledgment}
This research is supported in part by 
JSPS KAKENHI Grant Number JP.17H01729.

\end{document}